# Revisiting Monetarism: influence of Entropic Models.

Henry Daniel Vera Ramírez[1].

**Abstract**

This paper introduces an approach to *gas−like* models, from the concept of entropy, using the money stock data of two economic agents, in this case of two countries, which carry out market actions (trading) in two theoretical scenarios: in the absence of debt and with debt. The exercise deals exclusively with a no debt scenario and the data and results show that the bounded model generates low $P_{(m)}$, values that the results of the regressions between the two countries show an advantageous position of the stock country $m_i$ over the stock country $m_j$. About the rationale, it is found that these models can provide meaningful information regarding the behavior of monetary variables, -taking into account the different conceptual positions proposed in the manuscript-, using analogies derived from other fields of study ranging from molecules in rarefied gases or particles collisions, bringing the data to the interaction of economic actors.

**Key words**: Entropic models; Monetary Stock; Monetarism; *gas – like* Models. *JEL Classification*: **B16; E42; R15.**

## 1  Introduction.

Monetarism must be understood as a different phenomenon from that referred to by the tradition of the monetarist school. It refers to the acceptance of a rigid system of rules about the disciplined management of the currency which results in the recognition of the existence of an autonomous Central Bank, whose functions, in addition to issuing, focus on price control. Western economies have experienced the advantages of a monetary system in which the central bank plays a leading role.

In addition, its actions reflect ideas related to the sovereignty of the unit of account and its legal tender character. We can probably argue that this scheme has been instrumental in overcoming situations of complex inflation, in economic and monetary integration, and in maintaining the public's financial assets in the face of cyclical crises. However, it is also important to recognize that in systems where there has been a relationship of dependence between the metropolis and the periphery, the process of consolidation of a national currency is also subordinated to the overcoming of the *defects* produced by an inequality relationship.

---

[1] Universidade Nova de Lisboa; SENNOVA; h.ramirez@campus.fct.unl.pt

Nor a version of qualitative analysis that started from the *a priori* idea that the monetary system of the West comes as a *rescue* to a system of price exchange and issuance, which can generate not only setbacks in social aspects but in many other cases, affectations to the real economy and its variables of welfare and equality. Overcoming colonialism or the decolonial process also implies judging the effect of these processes of historical homogenization and proposing scenarios for overcoming current and future crises. For this reason, recognizing the existence of a process of social-economic-cultural domination also implies analyzing the arrival of monetary systems in global countries.

This manuscript attempts, in principle, to approach the set of theories that seek to explain the value of money in economic performance, with a brief review of the classical theories from antiquity through the Cambridge school to the school of Aftalion. In a second section, the concept of Entropy is explained about its application in the economic sciences, using explanatory schemes that relate prices to monetary resources.

In another section, a brief model of the application of methods using particle collision analogies is presented, using a model found in the study of gases and based on the entropic behavior of economic agents. In the fourth part, a brief implementation of the model based on the money stock of two (2) neighboring countries is discussed. Data from the application of the model are analyzed to obtain $P(m)$ values, related to a Gibbs distribution and values related to two (2) regressions between the two economic agents. The values of $P(m)$ and those of the regressions are extremely low, despite relatively high fit values.

## 2   Some Historical Aspects.

In principle, to refer to the process of progress that the acceptance of currency as a cultural signified implies, we must understand its transit, from the chartalism and metallist theories, the quantitative theories, the so-called Cambridge school, the Swedish and Austrian theories, the theory of performance and the so-called psychological school of Aftalion. From the chartalism conception, the currency has a value imposed on it by a supreme power. The currency being an instrument of exchange and being legal tender, it would have no other value than the one imposed on it by an authority which therefore allows monetary transactions (Oresme, 1956; Bresciani-Turroni, 1962).

The perspective of monetary chartalism is based on Aristotle, who already considered that the value of money came from the law, which was extended to the medieval period. However, in the Middle Ages, the metallist position came into contradiction with chartalism. Although chartalism guaranteed security in social relations, it began to lose acceptance in the Middle Age (Barré and Teulon, 1997; Baudin, 1947).

The metallist theories attributed to the coin an intrinsic value related to the *substance* embodied in the coin. The princes were dedicated to recognizing this value. One of the major proponents of the chartalism view of money was Knapp (1842-1926). Quantitative theories, on the other hand, are those that make the value of money dependent on its quantity, so that it is worth more when it is scarcer. The quantitative principle of the value of money appears in antiquity with Xenophon, considered the first quantitativist to whom the Roman jurisconsult *Iulius Paulus* refers as having a quantitative theory of the value of money (Digest, L. XVIII, quoted in Scott, 1932).

However, is Bernardo Davanzati (1529–1606), who first formulated the fundamental principle that would be the basis of quantitative theories, by affirming the existence of a mathematical relationship between gold and other goods. Subsequently, this quantitative principle was extended by Locke (1632-1704), Berkeley (1685-1753), Montesquieu (1689-1755), and David Hume (1711-1796). However, the explanation of the theory was taken to higher levels with the socalled English classics and mainly David Ricardo (1772-1823), who introduced into the old version a new element resulting from the evolution of monetary institutions, namely *paper money*.

Irving Fisher (1867-1947), establishes a relation between currency and prices which is expressed through the following equality relation $MV + M'V' = PT$, where M designates the monetary stock (whether in metal or paper), V, the velocity of circulation of money, M', is called the bank currency and V', the velocity of circulation of the bank currency; P, the level of prices and T, the volume of transactions (Fisher, 1892). Fisher's equation is merely tautological since it establishes a necessary equality between the number of payments and the prices of the goods obtained through these payments.

Changes in the quantity of currency in circulation lead to changes in the price level. These quantitative theories have some elements in favor of experience. It was only after the First World War that some monetary phenomena became clearer and did not necessarily conform to the quantitative explanation. Quantitative theories have a tendency principle. An increase or decrease in the money supply constitutes a high or low-price level factor, if there is no offsetting change in V, V' or T. It must also be understood that both V and V' can influence the general price level.

Quantitativists establish an automatic connection between a quantity of currency and the price level. This automatism is also the main defect of quantitative theories, which forget that economic phenomena are the consequence of human reactions, and therefore even changes in the quantity of currency cannot automatically and in a fixed proportion have an effect on the price level, since their influence must be exerted through the consciousness and will human beings, according to their reactions to monetary movements[2].

## 3 Performance Theory.

Since Cantillon (1680-1734) in the 18th century and Tooke (1774-1858) in the 19th century, the theory had already been formulated early, but it is an economist of the Austrian school, von Wiesner, already mentioned, in a presentation in 1901, in a paper at the assembly called *Verein fur Sozialpolitik¨* in Vienna, started from the principle that the study of the value of the currency should be done following the theory of the value of general goods. As a marginalist, he concludes that the value of a currency depends on the importance attributed to the last monetary unit, which does not come from its direct utility but from the purchasing power of the utility of the good that it allows acquiring. The last monetary unit is the last unit of the yield of each individual (Schumpeter, 1954).

Thus, the individual value of the currency decreases when this yield increases, and increases when it decreases. It could be thought that von Wieser's theory is only an extension of the quantitative theories. However, von Wieser offers a comprehensive explanation that the *quantitativists* do not. Price changes are not automatically determined by changes in the money supply, because they depend on subjective judgments (Hicks, 1967).

## 4 Cambridge and the *Robertson's* Equation.

The Cambridge School, through some of its representatives such as Marshall, Pigou, Robertson, and Keynes, outlined new quantitative interpretations of the problem of the value of money. The Cambridge School is quantitative to the extent that its psychologist allows it. In principle, Keynes had given his support to Fisher's construction, and he properly referred to equality as the refers to the Cambridge quantitative equation, which appeared devoid of any rigid automatism, subordinated to the idea of liquidity preference already expounded earlier by Marshall and Keynes.

For Robertson (1947), with P being the general price level of consumer goods, R the national yield, M the quantity of money, and K, a percentage of R that economic subjects wish to keep as liquid assets, we would have $P = M/KR$. According to Robertson, the general level of prices will depend on the quantity of M and the amount of credit that the economic subjects keep in liquid species. In this equation, Robertson conceives that M is related to the goods that constitute the yields of individuals, mostly consumer goods when it is true that also business people have M, taking into account the needs of their companies and the requirements of financial operations so that the formula could not represent the variations of all prices and consequently the variations of the value of M. Robertson, therefore, proposed another equation according to which *P = M/KT*. Now P' represents the price level of transactions and T the volume of transactions, while K' is the percentage of the volume of transactions that economic subjects wish to keep liquid. Pigou (1877-1959), also proposed a formula for changes equally close to the idea of *liquidity reserve*.

---

[2] However, the quantitative conception in the United States and France, has an enormous impact with Friedman (1912-2006) and Maurice Allis (1911-2010).

# 5 Keynesianism.

However, was Keynes who developed the idea of the study of the value of M and the concept of the complex unit of consumption or *purchasing power*, which corresponds to the set of typical articles that are the object of acquisition and consumption in a community. With the concept of the complex unit of consumption at the center, Keynes proposed the following formula: *N = P(K+ RK)*. In this equation, N represents M in circulation plus bank reserves. P is the overall price of the complex unit of consumption, K is the number of complex units of consumption for the acquisition of which M availabilities are conserved and K' is the number of complex units of consumption for the acquisition of which M banked are conserved. R is the ratio of bank reserves to deposits. The omission of any reference to the volume of transactions should be interpreted as being based on the assumption of a constant volume of transactions. For Keynes, the general price level will vary according to the increase or decrease in the money supply, holding other values constant; but it will also vary inversely when K and K' or R are altered (Keynes, 1939).

    The variations of K and K' depend on the habits of the communities and their structure, but R depends on the banking policy as in the case of the first Robertson equation. Keynes' equation only considers the prices of consumer goods, without consequently covering all the factors of variation in the value of M, influencing purchasing power. Keynes himself directed criticism of the formula presented. However, some advantages of his approach can be recognized, especially in substituting the price level abstraction for the price of certain goods. In the same way, the factors K and K´ introduced in the equation are dependent on the liquidity preference and are of the psychological root, very different from the factors V and V´ of Fisher's equation.

    In his general theory, Keynes adopts a simplified formula: *P = Y/O*. In this formula, Y stands for the M-pantries, and O for the volume of goods and services produced. Keynes concludes that only after a situation of full employment is reached does the increase in M produce an increase in the general price level. In effect, a shortage of M, characteristic of underemployment, with a high volume of goods produced but not traded. Also, due to the characteristics of

underemployment, the Keynesian formula seeks to keep prices relatively low. Thus, the possible effects of two movements of M on prices are limited to a situation of full employment (Keynes, 1936).

## 6 Swedish and Austrian.

Wicksell (1947) showed that prices vary due to the effect of variations in investments and that the volume of investments depends on the rate of the bank *juro* and the relation between it and the rate of return on the use of capital. Note the meeting point with Keynes, in the conclusion that it is only after a point of equilibrium has been reached in the monetary aggregates that price increases can be determined. Wicksell's monetary thesis was developed by the Swedish authors Eric Lindhal (1891-1960), Gunnar Myrdal (1898-1987), and Austrian von Hayek (1899-1992). Wicksell's earliest disciples made the price level partly dependent on anticipations of the future value attributed to goods, especially production goods. In this way, they introduced psychological elements into monetary theory. Wicksell himself had already revealed the influence of the psychological concerns of Von Wieser (1851-1926) and Bohn-Bawerk (1851-1914).

## 7 *Aftalion's* Psychological Theory.

This school is close to the yield theory. It tries to refine it, taking into account the phenomena that occurred during the First World War. For the school, the yield theory would only partially explain the behavior of the currency by applying the theory to the study of value. The value of a good depends on its rarity and utility. Therefore, von Wieser referred to these factors in the development of his exposition, which is to mention only the first one since the volume of the monetary yield represents only the rarity factor. The individual appreciation of M, as of any good, does not result from the satisfactions provided by the last extra unit, but from the satisfactions that everyone can expect from the use of the last unit. In this way, Aftalion, like several modern economists, introduces the element of *foresight* in the monetary theory.

When the currency is destined for acquisitions, hoarding will always be an expected utility that fixes its value. This utility will depend on the budget of a constant volume of transactions, not on its yield, but on the *forecast* of future variations in the purchasing power of the currency. Aftalion does not make it clear that in reality, the volume of transactions varies. Therefore, he considers changes in the supply of goods as a factor in the changes in the value of the currency and indicates their causes: fluctuations in quantities produced, in cost requirements; and forecasts as to these same costs, requirements, and quantities (Aftalion, 1950).

# 8    Entropy and Economics.

Figure (1) shows the relationship between money supply (M) and prices (P), through an entropy diagram of money stock-price level. Suppose the existence of a state of equilibrium *a*, where under ceteris-paribus conditions, the price level remains controlled and below one digit, with a certain quantity (M), and another scenario *b*, where due to reversible transformation conditions (which indicates that it is possible to return to the initial level of money supply and it is desirable to maintain the same price level), is represented by the area limited by the curve, which is given by:

$$\int_a^b PdM \quad (1)$$

In conditions of the natural equilibrium of the markets it corresponds to the total quantity of money stock absorbed by the system and the total quantity of the final money stock absorbed by the market in a second moment (Weston, 1959). If the process is reversible and cyclical, its representation would be a closed curve as in Figure (2) of the diagram. In this case, the area limited by the curve is nothing more than the amount of money stock absorbed by the system in the transformation process which represents the amount of market work translated into a unit of account. The areas of the diagram involve two other fundamental variables: productivity and technology:

$$P = P(p,t) \quad (2)$$

$$M = M(p,t) \quad (3)$$

That can be considered equivalent, while prices and money supply reflect the amount of work in the market. If the labor market, translated into the size of the labor force, has as a condition that its size is represented by the relationship between productivity and technology conditioned by the price level and the amount of money issued, we will have:

$$J = \frac{(p,t)}{(P,M)} = 1 \quad (4)$$

It should be equal to 1, under the ideal of making use of all resources. In the diagram, the processes of increase or decrease of money supply and prices are explained through parallel lines and coordinates (Fast, 1962; Chambadal, 1963). For a perfect market, the reversible isoquants are, in this diagram, congruent logarithmic curves, whose equation is:

$$M = CplnP + Cte \quad (5)$$

The slope of this curve at each point is given by:

$$\frac{\partial M}{\partial Pp} = \frac{Cp}{P} \tag{6}$$

where $C_p$, is the price level of the basic basket and that corresponds to greater pressure on the price level, while $C_te$, refers to a price concentration factor in areas due to high population factors or factors of social relevance. The isobar of a perfect market is logarithmic curves congruent with the equation:

$$M = CnBlnP + Cte \tag{7}$$

The slope of this curve at each point is given by:

$$\frac{\partial M}{\partial Tv} = \frac{Cv}{P} \tag{8}$$

is equal to the quotient of the price level that n corresponds to the $C_nB$ basic basket times the concentration factor and the general price level (Zemansky and Dittman, 1996).

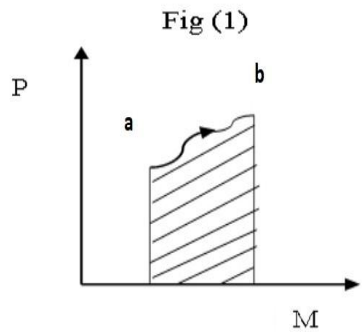

Figure 1: Behavior Prices and Monetary variables.

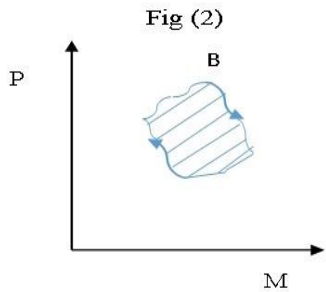

Figure 2: Behavior Prices and Monetary variables.

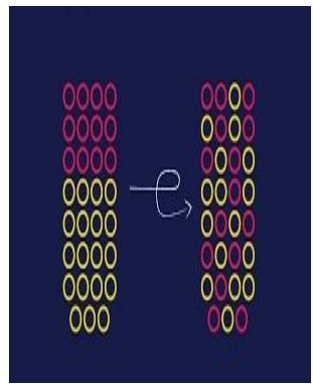

Figure 3: Example of Entropy.

# 9    Entropic Models.

To determine the relationship between monetary policies and the concept of Entropy, it is important to first define the physical meaning of the concept. Entropy can be defined as a state function of a thermodynamic system whose variation is closely related to the direction of its possible evolution. In an isolated system, only those transformations are possible in which Entropy acquires its maximum value. If the system is not isolated, it can carry out transformations in which Entropy grows. When thermodynamic equilibrium is reached, Entropy acquires its maximum value (Figure 3). If the system is not isolated, it can perform transformations in which Entropy decreases (Pippard, 1964).

In this case, for the Entropy of the universe (isolated system) to remain constant or increase, there must be an increase of Entropy in the interacting systems with the datum that is at least equivalent to the decrease of the datum. These facts are nothing more than an interpretation of the second principle of classical thermodynamics, enumerated by Clausius (1822-1888) and mathematically translated by the formula that bears his name. For an elementary transformation of Clausius's formula, we have:

$$dS \geq dQT \qquad (9)$$

where $dS$ is the variation of the Entropy of the system; $dQ$ is the amount of heat given up or received, and $T$ is the absolute temperature. In this equation, as well as those derived from it, the sign is = valid for reversible processes and the sign > for irreversible ones (case, for example, of the natural evolution of an isolated system). For a finite transformation between two states 1 and 2, a variation of the Entropy is given by:

$$S_2 - S_1 = \int_1^2 dS \geq \int_1^2 dQ/T \qquad (10)$$

If the transformation is cyclic, we have $S = K \log W$. In statistical mechanics, Entropy is defined by the Boltzmann (1844-1906) equation:

$$I$$
$$dQ/T \leq 0 \qquad (11)$$

with a logarithmic function of *thermodynamic probability W*, which represents the number of microstates corresponding to the macrostate. K is the Boltzmann constant. Thus W is so much larger when the system's disorder is larger. In nature, isolated systems tend to evolve in the direction of greater disorder. This is what happens when we have a container with insulated walls divided into two compartments, in each of which there is a gas. If the partition is removed, the two gases will spontaneously mix, the final state being more disordered than the initial one.

Therefore, Boltzmann considers that the Entropy grows, because the system passes from a state of lower probability to one of higher probability. It is very important that in statistical mechanics, the direction of the possible evolution of the system is determined by a law of probability. When two bodies are in contact at different temperatures, the heat passes from the hotter to the colder one, but there is a high probability that this will happen.

The concept of Entropy is contained in modern information theory or computer science. All information comes from messages. Suppose the probability of a message being sent is $p_i$. Our information about the message depends on the probability, where $I = I(p_i)$. If the probability $p_i$ is large, the information is small. The information then increases with $1/p_i$. According to information theory, information is related to the probability of the message multiplied by a logarithmic function called the Shannon[3], function $I = -C Log p_i$. C is a constant. It is therefore verified that the formal aspect of the Boltzmann equation of statistical mechanics has applicability in the use of information. Indeed, the more information there is about a system, the less disorder it has, and therefore the less entropy it has. Information can thus be *negentropic* (Brillouin, 2013).

---

[3] Claude Elwood Shannon (1916-2001).

However, there is a large body of work focusing more precisely on the relationship of the money supply to economic agents interacting in the market, under certain ceteris paribus conditions. The following model is one of them.

## 10    *Gas – like* **Models.**

About the free expansion of a gas, the following mental exercise can be used: first of all, imagine the existence of two (2) glass containers that are connected by a tube on which there is a key. A pump is then used to empty the containers of air. Then, if you keep the key closed, you will find that when you fill one of the containers with a rarefied gas (helium), the other one is empty (Silvestrini, 1985, p. 52).

If it is assumed that a gas is made up of a very small number of molecules and one gram of helium is taken, the number of molecules is about $10^{23}$. If atmospheric pressure is taken into account and a container volume of five (5) liters is found to contain one gram of helium (Silvestrini,1985, p. 52). Gas molecules can't be concentrated in a single compartment as a matter of pure probability. Maxwell (1831-1879) himself had referred to this process by stating that there couldn't be an imp to place each molecule in a specific compartment, contradicting *equistribution*. Suppose that two molecules, A and B, are arranged to enter the compartments. From a probabilistic point of view, A could unite in either the right or left compartment, as could molecule B. However, in the case of a shared probability, we could have both A in the first compartment and B in the other, or the reverse situation where B is in the first compartment and A occupies the other. This situation allows us to affirm that there is a probability of 2 to 1, between being located in an equidistributed way and concentrating in a single compartment (See table 1).

| Number of Molecules | Relationship probbabilities→ *Equidistribution* |
|---|---|
| 2 | 2 to 1 |
| 4 | 6 to 1 |
| 6 | 20 to 1 |
| 8 | 70 to 1 |
| 10 | 250 to 1 |
| 12 | 920 to 1 |
| 14 | 3.500 to 1 |
| 16 | 13.000 to 1 |

Table 1 *Equidistribution* Source: Silvestrini, 1985, p. 55.

Analogously, one can calculate the odds ratios in favor of equidistribution for a very large number of molecules. What can be observed, from the probability table, is that the probability increases about the number of molecules. If the number of molecules is as large as the number of molecules contained, the probability ratio is so high that the case where the molecules are all in one part is practically impossible. If one understands this phenomenon as a kind of measure of the disorder of a given configuration (e.g., half the number of molecules on either side), one could introduce a number $W^3$ representing how this configuration can be realized. The writing of this number and mainly of its power is the measure of disorder. This measure of disorder is entropy (See table 2).

| W | W Form $10^n$ | Represented by the exponent of W |
|---|---|---|
| 10 | $10^2$ | 2 |
| 1.000 | $10^4$ | 3 |
| 10.000 | $10^4$ | 4 |
| 1.000.000 | $10^6$ | 6 |
| 1.000.000.000 | $10^9$ | 9 |

Table 2. *Exponent of W*. Source: Silvestrini, 1985, p. 55.

It could be said that the configuration of a system consisting of many elements has high entropy when it can be realized in many possible ways, i.e. when the configuration is the most probable. It will have low entropy when it is less probable. From this, it is quite reasonable to infer that a system consisting of a very large number of components tends to evolve spontaneously towards situations of maximum entropy.

The study of the transfer between pairs that rationally trade money and the result of a statistical distribution of money is a fundamental element of analysis in modern economics. Econophysics has initiated a series of studies in which research on the theory of money is somewhat fundamental. The influence of money has been an extremely long work in which the dynamics of money are studied and associated on the one hand with a probability distribution of money between agents, which can only be obtained recently and from a numerical point of view under the possibility of theoretically investigating analogous models with, for example, the behavior of particles (Sinha, Chatterjee, Chakraborti, Chakarborti, 2011, p. 138).

---

[4] Number of ways in which a configuration can be obtained.

If the analogy of the two particles and their collision is followed, from the point of view of the outcome of the change in their individual kinetic energy (or momentum), models of money income exchange can be based on this assumption, allowing the interaction between agents to be analyzed. If there are two agents chosen randomly by some defined mechanism, it can be assumed that the exchange process will not depend on previous exchanges and their dynamics but will follow a Markovian-type process:

$$\begin{pmatrix} m_i(t+1) \\ m_j(t+1) \end{pmatrix} = M \begin{pmatrix} m_i(t) \\ m_j(t) \end{pmatrix} \qquad (12)$$

Where $m_i(t)$ is the income of agent $i$ at time $t$, and the collision matrix $M$ defines the exchange mechanism. In this classical model, it can be considered that in a closed economy where the total of money $M$, and the total of agents $N$, is complete, it corresponds to a situation in which there is no migration or production and the economic activity is to trade currency. Each agent $i$, whether individual or corporate, owns a quantity of money $m_i(t)$, at time $t$. In some exchanges, a pair of agents $i$ and $j$ exchange their money, for which all of it is locally conserved and none ends up with negative balances. That is, $m_i(t) \geq 0$, i.e. there can be no debt mechanism. This results in:

$$m_i(t+1) = m_i(t) + \Delta m; m_j(t+1) = m_j(t) - \Delta m \qquad (13)$$

and if you follow local conservation, you have:

$$m_i(t) + m_j(t) = m_i(t+1) + m_j(t+1) \qquad (14)$$

The time ($t$) changes for each time unit step in the transaction. The simplest model considers that a random fraction of the total money must be shared:

$$\Delta m = \epsilon_{ij}[m_i(t) + m_j(t)] - m_i(t) \qquad (15)$$

Where $\epsilon_{ij}$ is a random fraction ($0 \leq \epsilon_{ij} \leq 1$) changing time or trading. The steady-state ($t \to \infty$) distribution of money is a Gibbs distribution. For example:

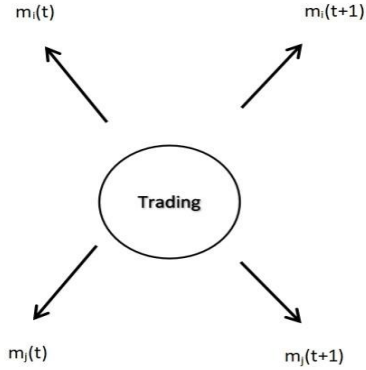

Figure 4: Schematic diagram of the trading process Source: Sinha, Chatterjee, Chakraborti, Chakarborti, 2011, p. 139.

$$P(m) = (1/T)exp(-m/T); T = M/N \qquad (16)$$

The Figure (4) shows the behavior of the agents during the exchange in each period:

Agents $i$ and $j$ redistribute their money in the market $m_i(t)$ and $m_j(t)$, this being the amount of their money before generating transactions. Changes $m_i(t+1)$ and $m_j(t+1)$ are generated after the transaction flow. To this extent, it is irrelevant how the uniformity justifies initial distribution. This means that in the case of a Gibbs distribution most economic agents have too much money. This is related to the conservation of money from the point of view of the additivity of entropy:

$$P(m_1)P(m_2) = P(m_1 + m_2) \qquad (17)$$

The result is in line with reality, since large variations in transaction flows and a possible coupling of agents who can only transact with their neighbors, as in the case of the proposed example, generate few and insignificant changes in the distribution. Some other variations, such as randomly shared shares with an amount of $2m_2$, when $(m_1 + m_2)$ when $m_1 \geq m_2$ and taking into account the classical idea of a low level of trade, can generate a drastic event, and this is that all the money in the market derives to a single agent, while the other becomes poorer.

If one of the economic agents takes on debt, things in the model undergo major changes. From the point of view of the individual, debt must be seen as a negative balance of money. Therefore, as the agent borrows money from the bank, the balance of $M$ increases, but at the expense of a debt or obligation $D$, which must be repaid and which generates a negative internal balance. The agent's total money will then be $M_b = M - D$, remaining stable. If one then takes into account a relaxation of the chaining condition of $m_i \geq 0$, $P(m)$ will never stabilize and will remain expanding towards infinity $m \to +\infty$ and $m \to -\infty$, according to Gauss. Total money is conserved and some agents get more wealth at the expense of others who have generated debt. Thus $M = M_b + D$.

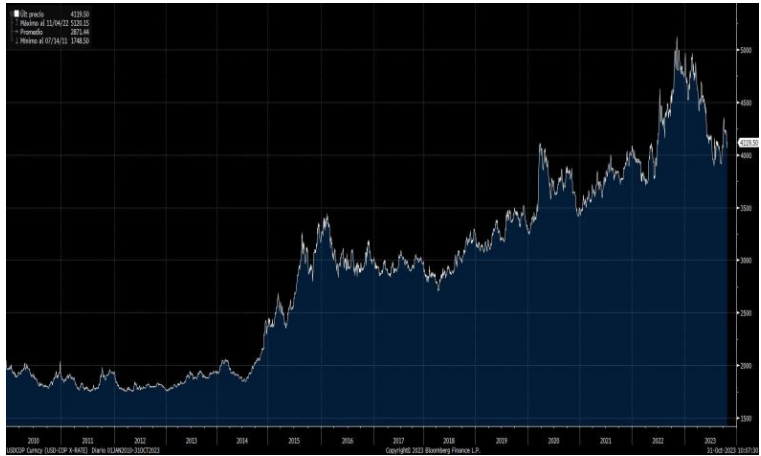

Figure 5: US COP Currency. Source: Bloomberg

Xi *et al.* (2005) restricted the total debt of agents participating in the money market. Generally, banks set aside a fraction of $R$ of their money deposited in bank accounts in such a way as to guarantee the permanence of $1-R$, so that it can be lent out. If the initial amount of money in the system is $M_b$, then, when repeated lending and borrowing takes place, the amount of money available to agents increases by $M = M_b/R$, where $1/R$ is the money multiplier and this extra money comes from the increase in the total debt of the system.

The total value of debt is $D = M_b/R - Mb$ and is bounded by the factor $R$. For a maximum level of debt, the amount for a positive balance is $(M_b/R)$ and the negative balance corresponds to $(M_b(1-R)/R)$, money circulating among agents in the system. The distribution of positive and negative balances is, together with their exponentials and with two different *temperatures*, corresponding to $T_+ = M_b/R$ and $T_- = M_b(1-R)/R$.

## 11   Results.

In Figures (4) and (5), we can see the behavior of the exchange rate of Colombia and Peru since 2010.

The money supply is the sum of currency outside banks; demand deposits other than those of the central government; time, savings, and foreign currency deposits of resident sectors other than the central government; bank and travelers' cheques; and other collateral such as certificates of deposit and negotiable instruments. Two economic agents were taken, corresponding in this case to two countries trading from their base money stock. Although the model was initially focused on individual and corporate economic agents, -under very conditional assumptions-, it is considered that extrapolation on data relatively close to reality can constitute an economic policy instrument for decision-making. The values of the monetary stock base have been taken to assume that this value corresponds to the values of $m_i$ and $m_j$.

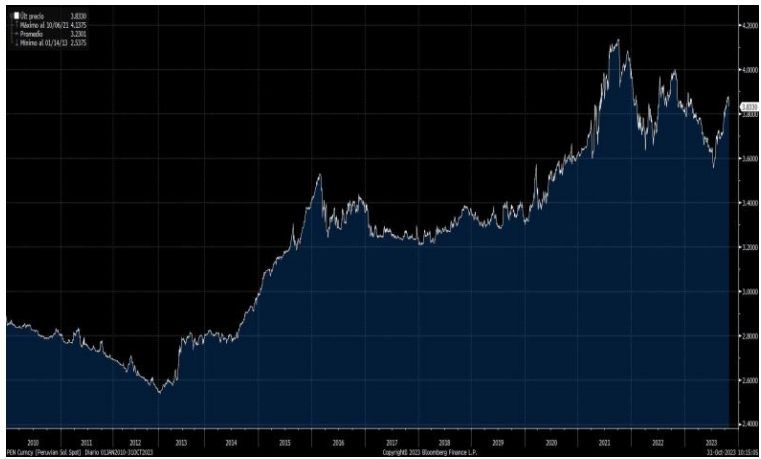

Figure 6: PEN Currency (Sol Peruvian Spot). Source: Bloomberg

In addition, we have the values of $M$, corresponding to the total sum of the monetary stocks of both countries. From the differentials, we obtain the $\Delta m$ values, which refer to the value of the difference between the individual money stocks $m_i$ and $m_j$ and the total stocks from period to period ($t$) and ($t+1$). In principle, the variables were operationalized as follows.

$M$= Sum of the monetary stocks of both agents, in this case of both countries. $m_i$ = Monetary stock (Colombia). $m_j$ = Monetary Stock (Peru).
$T$= Monetary stocks of both agents/number of agents.
$N$ = number of agents.
$\Delta m$= values of the difference between the individual money stocks in ($t$) and ($t+1$). In this case, the ratios of the money supply of each agent (country) to the total money supply for each period ($t$) and ($t+1$) were taken. $\epsilon_{ij}$= Random fraction.

Consequently, we have carried out an exercise that attempts to use the gas-like model, with two agents that carry out monetary transactions from a commercial point of view, with data on the money supply of the countries of Colombia[5] and Peru[6], from the year 2010-2019, according to data reported by the International Monetary Fund.

| Year | Colombia | Peru |
| --- | --- | --- |
| 2010 | 48.985.981.706 | 41.801.150.543 |
| 2011 | 58.256.101.520 | 46.707.014.697 |
| 2012 | 67.618.782.503 | 53.719.975.841 |
| 2013 | 76.905.990.533 | 62.572.474.316 |
| 2014 | 83.974.341.189 | 66.492.714.938 |
| 2015 | 93.565.102.388 | 75.394.378.147 |
| 2016 | 100.289.315.363 | 77.472.235.058 |
| 2017 | 107.098.732.080 | 84.701.003.507 |
| 2018 | 112.568.385.201 | 89.303.081.358 |
| 2019 | 122.532.476.451 | 96.731.613.884 |

Table 3. *Monetary Stock*. Source: International Monetary Fund, Report, 2020.

The Figures (7) and (8), line graph and a bar chart comparing the behavior of the money supply for both countries and the values :

| Year | Δ$m$ | M |
| --- | --- | --- |
| 2010 | - | 90787132249 |
| 2011 | 14.175.983.968 | 104.963.116.217 |
| 2012 | 16.375.642.127 | 121.338.758.344 |
| 2013 | 18.139.706.505 | 139.478.464.849 |
| 2014 | 10.988.591.278 | 150.467.056.127 |
| 2015 | 18.492.424.408 | 168.959.480.536 |
| 2016 | 8.802.069.885 | 177.761.550.420 |
| 2017 | 14.038.185.167 | 191.799.735.587 |
| 2018 | 10.071.730.972 | 201.871.466.559 |
| 2019 | 17.392.623.776 | 219.264.090.335 |

Table 4. Δ$m$ Source: International Monetary Fund, Report, 2020.

---

[5] A reference exchange rate was used with the last period with a value of 4213.4299, according to the Colombian representative market rate. This undoubtedly generates an exercise far from reality, since the exchange rate should refer to the weighted average of each year. However, in the case of the exposition of the case from the explanatory point of view, this does not distort the explanation of the model.

[6] An exchange rate of 3.8645587.

The behavior of the money supply of the variables is expressed using an exponential function, which includes the value T, corresponding to the total money supply for each period over the number of agents, $T = M/N$:

| $T = M/N$ | $(-m/T)$ | $(1/T)$ |
|---|---|---|
| 45.393.566.124 | -1.07914 | 0.0000000000220296 |
| 52.481.558.109 | -1.11003 | 0.0000000000190543 |
| 60.669.379.172 | -1.11455 | 0.0000000000164828 |
| 69.739.232.425 | -1.10277 | 0.0000000000143391 |
| 75.233.528.064 | -1.11618 | 0.0000000000132919 |
| 84.479.740.268 | -1.10754 | 0.0000000000118372 |
| 88.880.775.210 | -1.12836 | 0.0000000000112510 |
| 95.899.867.794 | -1.11678 | 0.0000000000104275 |
| 100.935.733.280 | -1.11525 | 0.0000000000099073 |
| 109.632.045.167 | -1.11767 | 0.0000000000091214 |

Table 5. ($1/T$) Source: International Monetary Fund, Report, 2020.

The reciprocal values of $T$ are extremely small. However, they were not expressed as negative powers to make them easier to understand. The values of the exponents of the values of $-m/T$ are then obtained and multiplied by the reciprocal of $T$.

| $exp(-m/T)$ | $(1/T)exp(-m/T)$ |
|---|---|
| 0.339887937 | 0.000000000007488 |
| 0.32954909 | 0.000000000006279 |
| 0.328064356 | 0.000000000005407 |
| 0.331951938 | 0.000000000004760 |
| 0.327527775 | 0.000000000004353 |
| 0.330369067 | 0.000000000003911 |
| 0.32356418 | 0.000000000003640 |
| 0.327333209 | 0.000000000003413 |
| 0.327833932 | 0.000000000003248 |
| 0.327040828 | 0.000000000002983 |

Table 6. ($1/T)exp(-m/T)$ Source: International Monetary Fund, Report, 2020.

The value of $P(m)$ for the total data is extremely small with a value of $4.54831*10^{-12}$.

Figure (9) shows the results of the regressions for each of the countries, according to the data from their money supply, for the data from 2010 to 2019. On the other hand, we found in Figure (8) the fraction of $m_i$ and $m_j$ in relation with total monetary stock:

| Year | $\Delta m = \epsilon_{ij}[m_i(t) + m_j(t)]m_i(t)(Colombia)$ | $\Delta m = \epsilon_{ij}[m_j(t) + m_i(t)]m_j(t)(Peru)$ |
|---|---|---|
| 2010 | 19.246.518.128,96070 | 26.431.349.291,99240 |
| 2011 | 20.783.922.015,38820 | 46.508.992.805,99760 |
| 2012 | 23.783.297.634,97290 | 54.057.746.424,09080 |
| 2013 | 28.071.104.355,14550 | 60.544.327.076,70800 |
| 2014 | 29.383.715.303,61430 | 57.853.932.833,10630 |
| 2015 | 33.643.050.027,06590 | 70.306.198.676,66040 |
| 2016 | 33.764.034.970,54150 | 65.383.185.160,39480 |
| 2017 | 37.404.952.478,47810 | 73.840.866.219,15770 |
| 2018 | 39.505.535.259,73060 | 72.842.570.074,39840 |
| 2019 | 42.674.589.853,20590 | 85.868.076.196,20560 |

Table 7. $\Delta m$ *Colombia Peru* Source: International Monetary Fund, Report, 2020.

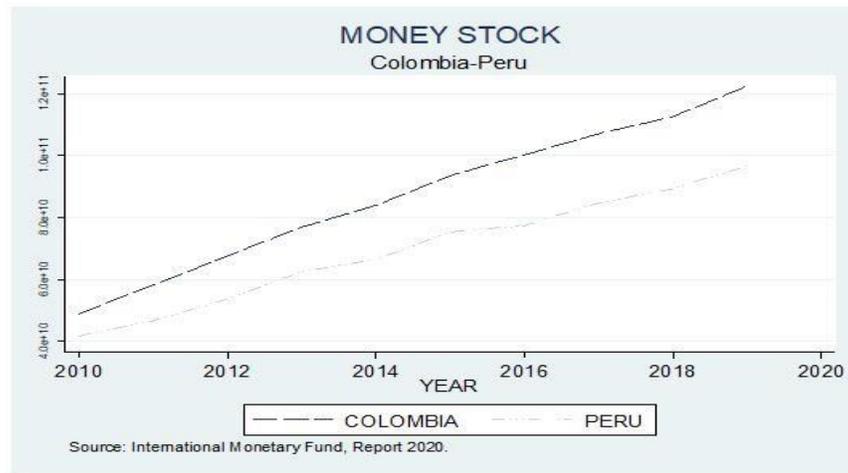

Figure 7: Relationship monetary stock Colombia-Peru. Source: Own construction.

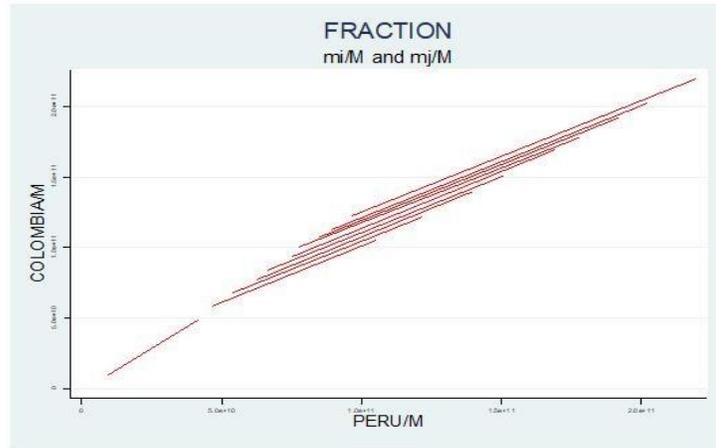

Figure 8: Fraction *m_i* and *m_j*. Source: Own construction.

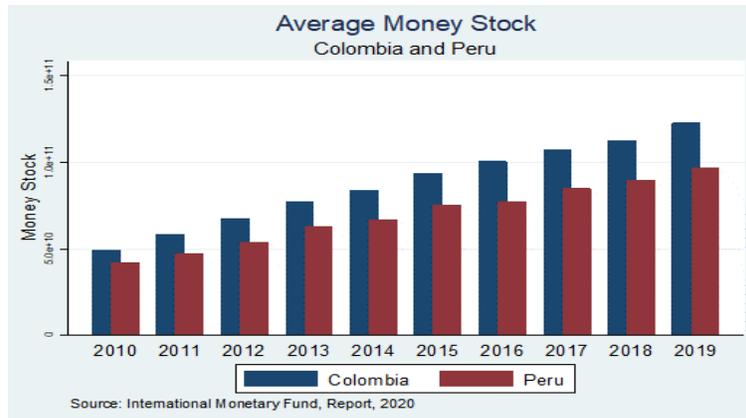

Figure 9: Relationship bar-graph monetary stock Colombia-Peru. Source: Own construction.

The regression values for the case of country $m_j$ concerning country $m_i$, show a $R$ value of 0.9968 with a $R - adjusted$ of 0.9964; the coefficient of the regression that indicates the variation of the monetary stock of Peru about that of Colombia, presents a value of 0.7582851, which indicates that a variation of one unit of monetary stock in Colombia, will generate an increase in the monetary stock of Peru of 0.7582851 USD. The value of $P > t$ is 0.0151948, well below 5%, and the value of the Peruvian money stock that does not depend on the variation of the Colombian stock is $3.38 * 10^9$ USD.

The regression values for the case of country $m_i$ with respect to country $m_j$, show a coefficient of the regression that indicates the variation of the monetary stock of Colombia in relation to that of Peru, $R$ of 1.314542, which indicates that a variation of one unit of monetary stock in Peru, will generate an increase in the monetary mass of Peru of 1.314542 USD. The value of P>t is well below 5% and the value of the Peruvian money stock that does not depend on the change in the Colombian stock is a negative value $-4.17 * 10^9$ USD.

## 12 Conclusions and Recommendations.

The results of the application of Gas-Like Models, using the data from two countries, showed an interesting exercise-like issue for developing economic politics in relation to the influence of monetary stock. In principle, that data and the outcomes show similar behavior in the monetary stock for every country with low values of $P(m)$, due to $m_i$ being greater than $m_j$, if we attended to the exchange rate. However, the use of the fraction $\epsilon_{ij}$ was arbitrary, due to the use of the fraction-like proportion of the total amount of the stock of $m_i$ and $m_j$ in relation to the total monetary stock.

The calculation exercises from the different tables suggest that in a scenario of no-debt the behavior of the monetary stock of two economic agents, could be analyzed like the behavior of the relationship between molecules in the gas and the collision of particles. The model does not explore the debt scenario but explores the probability of this situation in the background theory.

Entropic models and monetary economics are not directly related in the traditional sense, but they may have some indirect connections in the context of economic theory and decision-making. Some topics of research are:

*Information theory and economics*: Entropy is a fundamental concept in information theory, which is used to measure the uncertainty or lack of information in a system. In the field of economics, especially financial economics and decision-making, information theory can be relevant. For example, portfolio theory uses concepts such as diversification to reduce uncertainty and improve investment decision-making

*Behavioral Economics*: Monetary economics can be affected by human behavior, which is often irrational and influenced by cognitive biases. Entropy and related concepts such as asymmetric information and constrained decision making may play a role in understanding how economic agents process information and make decisions in a monetary environment.

*Prediction and volatility models*: Entropy and other information theory concepts are also used in modeling financial volatility and predicting movements in money and financial markets. Stochastic volatility models, such as the GARCH (Generalized Autoregressive Conditional Heteroskedasticity) model, incorporate entropy concepts to assess uncertainty and variability in the prices of financial assets.

In summary, although there is no direct relationship between entropy models and monetary economics, concepts related to entropy and information theory may have applications in economics, especially in understanding decision-making, financial volatility, and the management of uncertainty in monetary and financial markets. However, these connections are often more indirect and apply to specific areas within the economy.

# 13 References.

[1]  Aftalion, A. (1950). Monnaire, Prixet Change. Paris.

[2]  Baudin, L. (1947). La Monnaie et la Formation des Prix. Recueil Sirey. Paris.

[3]  Barré, R and Teulon, F. (1997). Economie Politique. PUF. Paris.´

[4]  Bresciani-Turroni, C. (1962). Corso di Economia Pol´ıtica. Giuffré. Vol II. Milan.

[5]  Brillouin, L. (2013). Science and Information Theory. Dover Publication
[6] Chambadal, P. (1963). Evolution et applications du concept d´ Entropie. DUNOD. Paris.

[7]  Fast, J. D. (1962). Entropy the Significance of the Concept of Entropy and its Applications in Science and Technology. Second Edition Revised. McGraw Hill.

[8]  Fisher, I. (1892). Mathematical investigations in the Theory of Value and Prices. Academy Arts and Science. Connecticut.

[9]  Hicks, J. (1967). Critical Essays in Monetary Theory. Oxford.

[10] Keynes, J. (1939). A treatise on Money. MacMillan and Co. London.

[11] Keynes, J. (1936). The General Theory of Employment, Interest and Money.MacMillan and Co. London.

[12] Oresme, N. (1956). The De Moneta. Thomas Nelson and Sons Ltd. London.

[13] Pippard, A. (1964). The Elements of Classical Thermodynamics. Cambridge University Press. London.

[14] Robertson, D. H. (1947). Money. Cambridge University Press/Nisbet Co. London.

[15] Schumpeter, J. (1954). History of Economic Analysis. New York.Money.

[16] Scott, S. P. (1932). The Enactments of Justinian. *The Digest or Pandects* Book L. The Civil Law, XI, Cincinnati.
Recovered from: https://droitromain.univgrenoble-alpes.fr/Anglica/D50$_S$cott.htm

[17] Silvestrini, V. (1985). Che cos'e l'entropia. Editori Riuniti.

[18] Sinha, S. Chatterjee, A. Chakraborti, A. Chakraborti, B. (2011). Econophysics: An introduction. WILEY-VCH.

[19] Weston, F. (1959). Teoría Cinética dos Gases e Mecánica Estadística. Reverté. Barcelona.

[20] Wicksell, K. (1947). Lecciones de Economía Política. M. Aguilar.

[21] Xi, N. Ding, N. and Wang, Y. (2005). How the required reserve ratio affects the distribution and velocity of money. Physica A: Statistical Mechanics and its Applications, vol. 357, issue 3, pp. 543-555. DOI: 10.1016/j.physa.2005.04.014

[22] Zemansky, M. and Dittman, R. (1996). Heat and Thermodynamics. McGraw-Hill College.

# 14  Appendix.

```
regress PERU COLOMBIA

      Source |       SS           df       MS
-------------+------------------------------------
       Model |  3.0461e+21         1    3.0461e+21
    Residual |  9.7850e+18         8    1.2231e+18
-------------+------------------------------------
       Total |  3.0559e+21         9    3.3954e+20

Number of obs =       10
F(  1,     8) =  2490.42
Prob > F      =   0.0000
R-squared     =   0.9968
Adj R-squared =   0.9964
Root MSE      =   1.1e+09

        PERU |      Coef.   Std. Err.      t    P>|t|     [95% Conf. Interval]
-------------+----------------------------------------------------------------
     COLOMBIA|   .7582851   .0151948    49.90   0.000     .7232457    .7933245
       _cons |   3.38e+09   1.37e+09     2.47   0.039     2.23e+08    6.54e+09
```

```
regress COLOMBIA PERU

      Source |       SS           df       MS
-------------+------------------------------------
       Model |  5.2806e+21         1    5.2806e+21
    Residual |  1.6963e+19         8    2.1204e+18
-------------+------------------------------------
       Total |  5.2976e+21         9    5.8862e+20

Number of obs =       10
F(  1,     8) =  2490.42
Prob > F      =   0.0000
R-squared     =   0.9968
Adj R-squared =   0.9964
Root MSE      =   1.5e+09

    COLOMBIA |      Coef.   Std. Err.      t    P>|t|     [95% Conf. Interval]
-------------+----------------------------------------------------------------
        PERU |   1.314542   .0263414    49.90   0.000     1.253799    1.375286
       _cons |  -4.17e+09   1.89e+09    -2.21   0.058    -8.52e+09    1.85e+08
```

Figure 10: Regressions. Source: Own construction.